\documentclass[aps,prd,preprint,floatfix,nofootinbib,superscriptaddress]{revtex4}

\usepackage{graphicx,epsfig}

\def\ie{{\it i.e.}}

\def\bra#1{\mathinner{\langle{#1}|}}
\def\ket#1{\mathinner{|{#1}\rangle}}

\def\psla{/\!\!\!p}

\newcommand{\GeV}{\unskip\,\mathrm{GeV}}
\newcommand{\MeV}{\unskip\,\mathrm{MeV}}

\newcommand{\Sigp}{\Sigma^+}
\newcommand{\Sigm}{\Sigma^-}
\newcommand{\mum}{\mu^-}
\newcommand{\mup}{\mu^+}

\newcommand{\Xim}{\Xi^-}
\newcommand{\beq}{\begin{equation}}
\newcommand{\eeq}{\end{equation}}
\newcommand{\bea}{\begin{eqnarray}}
\newcommand{\eea}{\end{eqnarray}}
\newcommand{\sigtopx}{\Sigma^+ \to p X^0}
\newcommand{\mmu}{m_\mu}
\newcommand{\mxo}{m_{X^0}}
\newcommand{\omgm}{\Omega^-}
\newcommand{\momg}{m_{\Omega^-}}
\newcommand{\mXim}{m_{\Xi^-}}

\newcommand{\gsim}
{\;\raisebox{-.3em}{$\stackrel{\displaystyle >}{\sim}$}\;}

\begin{document}

\begin{flushright}
OITS-771\\
hep-ph/0509081\\
\end{flushright}


\bibliographystyle{revtex}

\title{On the Possibility of a New Boson $X^0\,(214 \MeV)$ in
$\Sigp \to p\mu^+ \mu^-$} 




\author{N.G. Deshpande}
\affiliation{Institute for Theoretical Science, University of Oregon,
Eugene, OR 97403}

\author{G. Eilam}
\affiliation{Institute for Theoretical Science, University of Oregon,
Eugene, OR 97403}
\affiliation{Department of Physics, Technion-Israel Institute of
Technology, 32000 Haifa, Israel}

\author{J. Jiang}
\affiliation{Institute for Theoretical Science, University of Oregon,
Eugene, OR 97403}


\date{\today}

\begin{abstract}
We consider the existence of the state $X^0\,(214 \MeV)$ in $\Sigma^+
\to p \mu^+ \mu^-$ decay found by the HyperCP collaboration.  We
assume that a fundamental spin zero boson $X^0$ coupled to quarks
leads to flavor changing $s \to d X^0$ process.  We estimate the
scalar and pseudoscalar coupling constants by considering $\Sigma^+
\to p X^0$ and $K^+ \to \pi^+ X^0$ processes, and find that
pseudoscalar coupling dominates.  We then evaluate the branching
ratios for $K_L \to \pi^0 \pi^0 X^0$, $\pi^+ \pi^- X^0$ and $\Omega^- \to
\Xi^- X^0$ decays.  All these rates are found to be in the measurable
ranges.  We also comment on $X^0$ coupling to muons and constraints
from muon $g-2$.
\end{abstract}

\maketitle



Recent search for the rare decay $\Sigp \to p \mu^+ \mu^-$ by the HyperCP
collaboration~\cite{Park:2005ek} has found three events all clustered
at the dimuon mass of approximately $214 \MeV$.  The branching ratio
based on these three events for the process $\Sigp \to p \mu^+ \mu^-$
is quoted as $(8.6_{-5.4}^{+6.6}\pm5.5) \times 10^{-8}$.  The Standard
Model branching ratio for this process is dominated by long distance
contributions~\cite{He:2005yn}, which are estimated to give a
branching ratio between $1.6 \times 10^{-8}$ to $9.0 \times 10^{-8}$,
in agreement with the total rate, but not with the dimuon mass
distribution.  It is therefore intriguing to consider the possibility
that the dimuon events seen by HyperCP collaboration are due to a
hitherto unobserved new particle $X^0$ of mass $214 \MeV$.  Indeed,
the suggestion is made in Ref.~\cite{Park:2005ek} that if such a
particle exists, the branching ratio for the process $\Sigp \to p
X^0$, assuming that $X^0$ decays predominantly into $\mu^+ \mu^-$, is
\beq
{\rm BR}(\Sigp \to p X^0, X^0 \to \mu^+ \mu^-) = (3.1 _{-1.9}^{+2.4} \pm
1.5) \times 10^{-8}\,.
\label{eq:br1}
\eeq 
In this note we explore consequences of this assumption to other
rare decays, and suggest ways to either confirm or refute this
hypothesis.  We shall adopt the central value of the branching ratio
in Eq.~(\ref{eq:br1}) to obtain our estimates.  All our estimates are
subject to errors in Eq.~(\ref{eq:br1}), which are about $50\%$
statistical and $60\%$ systematic.

Our basic assumption is that there is a new spin zero boson $X^0$ of
mass $214 \MeV$ and its fundamental interaction is given by
\beq
{\mathcal L}_{int} = [\bar{d} (h_1 + h_2 \gamma_5) s X^0 + h.c.]+
\bar{\mu}(l_1 + i l_2 \gamma_5 ) \mu X^0\,.
\label{eq:lagr}
\eeq
There may be other interactions of $X^0$ with quarks and leptons, but
we do not make any {\it a priori} assumptions about these, leaving
them to be determined by future experiments.  We first derive
constraints on the flavor changing couplings $h_1$ and $h_2$ from the
decays $\Sigp \to p X^0$, and $K^+ \to \pi^+ X^0$.  We then calculated
the expected rates for $K_L \to \pi^+ \pi^- X^0$, $K_L \to \pi^0 \pi^0
X^0$ and $\omgm \to \Xim X^0$ .  All these rates are in the
experimentally accessible ranges.

To relate $\Sigp \to p X^0$ to the parameters in Eq.~(\ref{eq:lagr}),
we use matrix elements involved in hyperon semileptonic decays, and
take divergences on both sides. We use the standard hypotheses of CVC
and PCAC which are known to be excellent for the low energy processes.
We find
\beq
\bra{p} \bar{d} s \ket{\Sigp} = f_1 \frac{m_{\Sigp} - m_p}{m_s -
m_d} \overline{u}_p u_{\Sigp}\,,
\eeq
and
\beq
\bra{p} \bar{d} \gamma_5 s \ket{\Sigp} = g_1
\frac{m^2_{K^0}}{m^2_{K^0}-m^2_{X^0}} \frac{m_{\Sigp} + m_p}{m_s +
m_d}  \overline{u}_p \gamma_5 u_{\Sigp}\,,
\eeq
where $f_1$ and $g_1$ are vector and axial vector form factors.  For
quark masses, we use, $m_d = 7 \MeV$ and $m_s = 140 \MeV$ and for
proton and meson masses we use values from
Ref.~\cite{Eidelman:2004wy}.  We have related $g_3$ that occurs in the
axial current matrix element to $g_1$ using
PCAC~\cite{Donoghue:1992dd}.  Further, $|f_1| = 1$ from CVC and $|g_1|
= 0.33$ from $\Sigm$ hyperon decay~\cite{Eidelman:2004wy}.  Hence, the
$\Sigp \to p X^0$ decay width is given by~\cite{Donoghue:1992dd}
\beq
\Gamma(\Sigp \to p X^0) = \frac{q (E_p + m_p )}{4 \pi
m_{\Sigp}}(|A|^2 + |B|^2)\,,
\eeq
where
\bea
A &=& \frac{f_1 (m_{\Sigp} - m_p)}{m_s - m_d} h_1\,, \\
B &=& \frac{g_1 m^2_{K^0}}{m^2_{K^0} - m_{X^0}^2}\frac{m_{\Sigp}+m_p}{m_s
+ m_d}\left(\frac{E_p - m_p}{E_p + m_p}\right)^{\frac{1}{2}} h_2\,.
\eea
Here $q = \lambda^{1/2}(m_{\Sigp}^2,m_p^2,m_{X^0}^2)/(2 m_{\Sigp})$
with $\lambda(x,y,z) \equiv (x-y-z)^2-4 y z$, $q$ is the magnitude of
the final state momentum, and $E_p$ is the proton energy in the rest
frame of $\Sigp$.  Knowing the branching ratio of
$\sigtopx$~\cite{Park:2005ek}, we get
\beq
|h_1|^2 + 0.19 |h_2|^2 \simeq 9.14 \times 10^{-21}\,.
\label{eq:h1h2}
\eeq
We now obtain a constraint on $h_1 \bar{s} d X^0$ coupling by
considering the process $K^+ \to \pi^+ X^0$.  Since $K$ and $\pi$ are
both pseudoscalar, only the scalar coupling of $X^0$ is involved in
this process.  Again, taking the divergence of the vector current, we
can evaluate the width as 
\beq
\Gamma(K^+ \to \pi X^0) = \frac{1}{16 \pi m_{K^+}^3}
\lambda^{\frac{1}{2}}(m_{K^+}^2, m_\pi^2,
m_{X^0}^2) \left(\frac{m_{K^+}^2 -
m_\pi^2}{m_s - m_d} \right)^2  |h_1|^2\,. 
\eeq
Comparing this with the observed branching ratio for $K^+ \to \pi^+
\mup \mum$ of $8.1 \times 10^{-8}$~\cite{Eidelman:2004wy}, we obtain
the constraint
\beq
|h_1| < 7.4 \times 10^{-12}\,. 
\eeq
This bound can be further improved by using the branching ratio for
dimuons of invariant mass close to $214 \MeV$.  We thus see that $X^0$
coupling to $\bar{s} d$ is predominantly of the pseudoscalar type and
we have
\beq
|h_2| \gsim 3.6 \times 10^{-10}\,.
\eeq
We have examined the consequences of $X^0$ exchange in $K_L$-$K_S$
mass difference and the decay of $K_L \to \mu^+ \mu^-$.  With the
above value of $|h_2|$, the contribution of $X^0$ to the mass
difference $\Delta M_K$ (using vacuum saturation) is about five
orders of magnitude smaller than the experimental value.  Similarly,
using constraints on $X^0$ coupling to muons from $g-2$, discussed
later, we find $K_L \to \mu^+ \mu^-$ decay branching ratio to be
several orders of magnitude below the experimental value.  Thus, the
kaon mass difference and the $K_L \to \mu^+ \mu^-$ decay do not impose
useful constraints.

We now discuss the decays $K_L \to \pi^+ \pi^- X^0$ and $K_L \to \pi^0
\pi^0 X^0$ which are sensitive to the pseudoscalar coupling.  These
decays have been discussed by Ref.~\cite{Gorbunov:2000cz} in the
context of a chiral Lagrangian.  The approximation made in
Ref.~\cite{Gorbunov:2000cz} neglects momentum dependence of the matrix
element.  Instead we shall use experimental data on $K_{e4}$ decays to
estimate the decay rates.

In the decay $K^+(P+L) \to \pi^+(p_1) + \pi^-(p_2) + e^+(p_e) +
\nu_e(p_{\nu})$, the axial matrix element is given by~\cite{Wienke:1972vu}
\beq
m_K \bra{\pi^+ \pi^-} A_\mu \ket{K^+} = F P_\mu + G Q_\mu + R L_\mu
\eeq
where $P = p_1 + p_2$, $Q= p_1 - p_2$, $L= p_e + p_\nu$ and the
experimentally determined~\cite{Pislak:2001bf} form factors $F$ and
$G$ are found to be insensitive to variation in $\pi \pi$ invariant
mass, and we take the values to be $F = 5.832$ and $G=4.703$.  The
form factor $R$ is not accessible in the experiment.  However, a
reliable estimate is provided using current algebra and soft pion
approximation~\cite{Weinberg:1966kf}.  The value of $R$ for process
$K_L(P+L) \to \pi^+(p_1)+\pi^-(p_2)+X^0(L)$ is found to be
\beq
R = \frac{m_K}{f_\pi}\frac{(P+L)\cdot p_2}{(P+L)\cdot P}\,,
\eeq
where the definition of $P$ and $Q$ stay the same as above, while $L$
is now the momentum of $X^0$.  Using $SU(2)$ symmetry, and taking
divergence of the current, we have for the matrix element for $K_L$ decay
\beq
{\mathcal M} = \frac{\sqrt{2}{\rm Re}h_2 [F (P\cdot L) + G (Q\cdot L) + R L^2]}{m_K
(m_s + m_d)}
\eeq
We then integrated over the final state phase space and find the
branching ratios to be ${\rm BR}(K_L \to \pi^+ \pi^- X^0) = 1.54
\times 10^{-9}$ and ${\rm BR}(K_L \to \pi^0 \pi^0 X^0) = 8.02
\times 10^{-9}$.  The enhancement in the decay to $\pi^0$ is due to
the larger final state phase space.  Both these branching ratios are
large enough to be measured in $K_L$ decays. Compared to the rates
estimated using Ref.~\cite{Gorbunov:2000cz}, our results are about a
factor of $5$ smaller.

From the $h_2$ value, we can also predict the branching ratio of
$\omgm \to \Xim X^0$ decay.  We again consider the semileptonic
process $\omgm \to \Xi^0 + e^- + \bar{\nu}_e$~\cite{Finjord:1979pz}
and take the divergence of the axial current.  We find the matrix
element for $\omgm \to \Xim X^0$ is
\beq
{\mathcal M} = \frac{C_A h_2}{m_s + m_d} \bar{u}_{\Xim} k_\mu
u_{\omgm}^\mu\,, 
\eeq
where $C_A = -2.08$~\cite{Finjord:1979pz}, and $k_\mu$ is four
momentum of $X^0$.  Utilizing the projection operator for spin-$3/2$
fields
\beq
\Lambda_{\mu\nu}(p) = (\psla + M)\left(- g_{\mu\nu} + \frac{1}{3} \gamma_\mu
\gamma_\nu + \frac{2 p_\mu p_\nu}{3 M^2} - \frac{p_\mu \gamma_\nu -
p_\nu \gamma_\mu}{3 M} \right)\,,
\eeq
the $\omgm \to  \Xim X^0$ decay width is evaluated to be
\beq
\Gamma(\omgm \to  \Xim X^0) = \frac{|h_2|^2 C_A^2}{192 \pi \momg^5 (m_s
+ m_d)^2}
\lambda^{\frac{3}{2}}(\momg^2, \mXim^2, \mxo^2)[(\momg + \mXim)^2 - \mxo^2]\,,
\eeq
The branching ratio we get
is $2.1 \times 10^{-6}$, which is accessible at the HyperCP
experiment.

Assuming $X^0$ couples to $\bar{\mu}\mu$ as in Eq.~(\ref{eq:lagr}), it
will contribute to the muon $g-2$.  From Ref.~\cite{Jackiw:1972jz}, we
find
\beq
\Delta a_\mu = \frac{\mmu^2}{8 \pi^2}\int_0^1 x^2 d x \frac{l_1^2
(2-x) - l_2^2 x}{\mmu^2 x^2 + \mxo^2 (1-x)}\,.
\eeq
By requiring the $X^0$ contribution be smaller than the difference
between the experimental value and the theoretical prediction, \ie,
$\Delta a_\mu < 250 \times 10^{-11}$~\cite{Czarnecki:2005th}, and
assuming the $X^0$ is either a pure scalar $S^0$ or a pure
pseudoscalar $P^0$, we can constraint the couplings $l_1 < 8.6 \times
10^{-4}$ and $l_2 < 1.0 \times 10^{-3}$.  The decay widths of $S^0$
and $P^0$ to $\bar{\mu} \mu$ are given
\bea
\Gamma(S^0 \to \bar{\mu} \mu) &=& \frac{l_1^2 \mxo}{8 \pi} \left( 1-
\frac{4 \mmu^2}{\mxo^2} \right)^\frac{3}{2}\,, \\
\Gamma(P^0 \to \bar{\mu} \mu) &=& \frac{l_2^2 \mxo}{8 \pi} \left( 1-
\frac{4 \mmu^2}{\mxo^2} \right)^\frac{1}{2}\,.
\eea
Hence the derived decay widths are $\Gamma_S < 1.58 \times 10^{-11}
\GeV$ ($c\tau_S > 12 \mu m$) and $\Gamma_P < 1.6 \times 10^{-9} \GeV$
($c \tau_P > 0.16 \mu m$).  Unless the $\Gamma_S$ is more than an
order of magnitude smaller than the above limit, these lifetimes are
too short to be observed as displaced vertices.  If BR($X^0 \to \mu^+
\mu^-$) is less than $100\%$, then the total width would be larger and
the lifetime shorter.

Although we have assumed the branching ratio of $100\%$ to
$\mu^+\mu^-$, our predictions for $K_L \to \pi^+ \pi^- X^0$, $K_L \to
\pi^0 \pi^0 X^0$ and $\omgm \to \Xim X^0$ with subsequent decay $X^0
\to \mu^+ \mu^-$ do not change, if the branching ratio is smaller.
The reason is that an increase in the coupling constant $h_2$ compensates
exactly for the decrease in the branching ratio.  At present there does
not seem to be any evidence for the decays $X^0 \to e^+ e^-$, $X^0 \to
\gamma \gamma$, or $X^0 \to \nu \bar\nu$.  It would be interesting to
obtain experimental limits on these processes to further understand
the properties of $X^0$.

We have taken a purely phenomenological approach to the existence of a
particle $X^0$.  The possibility of a light, weakly interacting scalar
or pseudoscalar boson has been considered in the context of
supersymmetry in models of spontaneous supersymmetry
breaking~\cite{Ellis:1984kd,Bhattacharya:1988ey,Giudice:1998bp,Gorbunov:2000th}.
Specific effects in $K$ decays due to pseudoscalar bosons are
studied in Ref.~\cite{Gorbunov:2000cz} and the contributions to muon
$g-2$ are discussed in Ref.~\cite{Brignole:1999gf}.  Such particles can also
arise in theories of broken family symmetries and are known as
familons~\cite{Wilczek:1983wf}.  The coupling constants in such
theories are usually arbitrary and an analysis like ours is useful in
constraining the couplings.

In conclusion, we are unable to rule out the existence of $X^0$
particle based on present data.  We have suggested signals for this
particle to be observed in $K_L$ and $\omgm$ decays.  We believe the
branching ratios for the proposed processes are large enough to verify
existence of $X^0$ or to rule it out.  Confirmation of this boson
would be evidence of physics beyond the Standard Model.

Note added: As we were ready to submit our manuscript, we received a
communication from Xiao-Gang He that He, Tandean and Valencia had
submitted Ref.~\cite{He:2005we} to the arXiv, which studies some of
the same processes involving $X^0$.

\begin{acknowledgments}
We thank Xiao-Gang He for useful correspondence.  GE would like to
thank the Israel Science Foundation for partial support. This research
was supported in part by the U.S.~Department of Energy under Grants
No.~DE-FG02-96ER40969.
\end{acknowledgments}

\end{document}